\documentclass[10pt]{article}
\usepackage{graphics}
\usepackage{amsfonts}
\usepackage{amsmath}
\usepackage{amssymb}
             \let\p=\pi    \let\r=\rho
 \let\t=\tau    \let\ph=\varphi
\let\ps=\psi

\def\\{\hfill\break} \let\==\equiv

\let\0=\noindent
\def\bra#1{{\langle#1|}}\def\ket#1{{|#1\rangle}}
\def\media#1{{\langle#1\rangle}}

\def\nn{\nonumber}

\def\qed{\hfill\raise1pt\hbox{\vrule height5pt width5pt depth0pt}}
\let\up=\uparrow      
\let\bu=\bullet
\def\be{\begin{equation}}
\def\ee{\end{equation}}
\def\bea{\begin{eqnarray}}\def\eea{\end{eqnarray}}
\begin{document}
%------------------------------------------------------------------------------
\markright{Crystallizing Universe...}
%------------------------------------------------------------------------------
\title{Time and Spacetime: The Crystallizing Block Universe}
\author{George F. R. Ellis$^*$ and Tony Rothman$^{*\dagger}$
\\[2mm]
{\small\it \thanks{george.ellis@uct.ac.za}}~ \it $^{\dag}$University of Cape Town, \\
\it Rondebosch 7701, Cape, South Africa\\
{\small\it\thanks{trothman@princeton.edu}}~ \it Princeton University, \\
\it Princeton, NJ 08544}
\date{{\small   \LaTeX-ed \today}}
%-----------------------------------------------------------------------------
\maketitle
%-----------------------------------------------------------------------------

\begin{abstract}
The nature of the future is completely different from the nature of the past.
When quantum effects are significant, the future shows all the signs of quantum weirdness,
including duality, uncertainty, and entanglement. With the passage of time, after the time-irreversible
process of state-vector reduction has taken place, the past emerges, with the previous quantum
uncertainty replaced by the classical certainty of definite particle identities and states. The
present time is where this transition largely takes place, but the process does not take place uniformly:
Evidence from delayed choice and related experiments shows that isolated patches of quantum
indeterminacy remain, and that their transition from probability to certainty only takes place later.
Thus, when quantum effects are significant,   the picture of a classical Evolving Block Universe (`EBU')
cedes place to one of a Crystallizing Block Universe (`CBU'), which reflects this quantum transition from
indeterminacy to certainty, while nevertheless resembling  the EBU on large enough scales.
 \vspace*{5mm}
 %\noindent PACS: ** GE: omit, being phased out! **
\\ Keywords: Space-time, quantum uncertainty, arrow of time, block universe.
\end{abstract}

%-----------------------------------------------------------------------------
\section{The Nature of Space Time: %Classical and Quantum
Emergent universes}
\setcounter{equation}{0}\label{sec1}
%-----------------------------------------------------------------------------
\baselineskip 8mm

According to the  spacetime view
associated  with both special  and  general
relativity, time, in a real sense, is little more than an illusion.   Given
data at an arbitrary instant, we assume that everything occurring at a later or earlier time
is  determined, evolved according to time-reversible local
physics.  Consequently,  nothing is, or can be, special about any particular moment.
The standard spacetime diagrams used in relativity enforce such a
view: no
special status is accorded  to the present  and indeed the
``now" is not usually even denoted on the diagram. Rather, all
possible ``presents" are simultaneously represented on an equal footing.
In a few  instances, cosmology takes into account time-irreversible physics
(for example nucleosynthesis in the early universe), but the
notion of the present as a special time remains absent.

Such a view can be formalized in the idea of a block universe
\cite{Mel98,Sav01,Dav02}: space and time are represented as merged
into an unchanging spacetime entity, with no particular space
sections identified as the present and no evolution of spacetime
taking place. The universe just \emph{is}: a fixed spacetime block.
In effect such a representation embodies the idea that time is an
illusion: time does not ``roll on" in this picture. All past and
future times are equally present, and the present ``now" is just one
of an infinite number.  Price \cite{Pri96} and Barbour \cite{Bar99}
in particular advocate such a position.\footnote{And see the debates
about the idea in the fqxi website for their \emph{Nature of Time}
essay competition, to be found at
http://www.fqxi.org/community/forum/category/10; many of the essays
there support the Block Universe idea.} Underlying the idea,
 as emphasized by Barbour, is that time-reversible Hamiltonian dynamics provides the foundation
 for
physical theory.

By contrast to this view, in a previous paper \cite{Ell06} one of us
has argued that the true nature of spacetime is best represented as
an Emergent Block Universe (EBU), which adequately represents the
differences between the past, present, and future, and which depicts
the change from the potentialities of the future to the determinate
nature of the past. The main feature of the EBU is an indefinite
spacetime, which grows and incorporates ever more events,
``concretizing" as time evolves along each world line. That paper,
however, was based on a classical view of physics. The present work
extends the Emergent Block Universe view to one designed to take
account of the quantum nature of physics at microscopic scales. We
represent the effects of ``quantum weirdness" through  a
``Crystallizing Block Universe" (CBU), where ``the present" is
effectively the transition region in which quantum uncertainty
changes to classical definiteness. Such a crystallization, however,
does not take place simultaneously, as it does in the simple
classical picture. Quantum physics appears to allow some degree of
influence of the present on the past, as indicated by such effects
as Wheeler's delayed choice experiments and Scully's quantum eraser
(see the summaries of these effects in \cite{AhaRoh05,GreZaj06}).
Our CBU picture adequately reflects such effects by distinguishing
the transitional events where uncertainty changes to certainty,
which may in some cases be delayed till after the apparent ``present
time."

The CBU is of course a riposte to those proposing an unchanging
Block Universe picture as an adequate representation of spacetime
structure on all scales. In our view that picture (based on
time-reversible Hamiltonian dynamics) does not represent adequately
either the macroscopic arrow of time, nor the unpredictable and
time-irreversible quantum measurement process. We first examine the
classical case, and then explore how it must be modified by
considerations arising from quantum physics.

%-----------------------------------------------------------------------------
\section{The Emerging Block Universe: The Classical Picture}
\setcounter{equation}{0}\label{sec2}
%-----------------------------------------------------------------------------

How do we envisage spacetime and the objects in it as time unrolls?
To motivate the classical Evolving Block Universe (`EBU') model of
reality, consider the following scenario \cite{Ell06}: A massive
object has rocket engines attached at each end that allow it to move
either left or right.  The engines  are fired alternately by a
computer, which  produces firing intervals and burn lengths based on
the random decays of a radioactive element. These signals
originating from the decaying element thus select  the actual
spacetime path of the object from the set of all possible paths. Due
to the quantum uncertainty in the radioactive decays, the realized
path is not determined by initial data at any previous time (see
Figure 1).\\

%---------------------------------------------------------------------------------------------------------

\begin{figure}[htb] \vbox{\hfil\scalebox{.5}
{\includegraphics{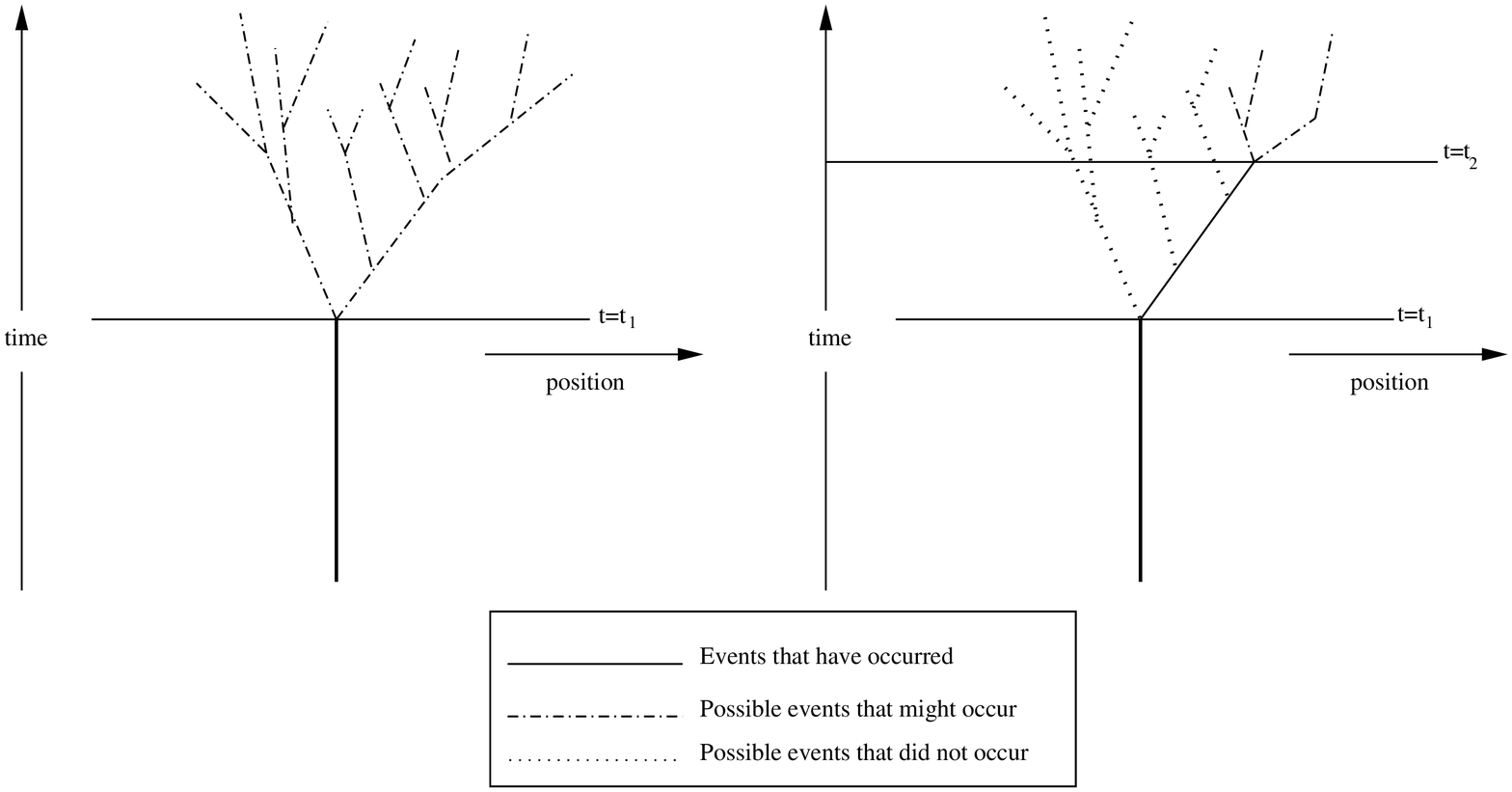}}\hfil} {\caption{\footnotesize{Motion of a particle world line controlled in
a random way, so that what happens is determined as it happens. On
the left events are determined till time $t_1$ but not thereafter;
on the right, events are determined till time $t_2 > t_1$, but not
thereafter.
 \label{fig1}}}}
\end{figure}

%-----------------------------------------------------------------------------------------------------------

Because the objects are massive and hence produce spacetime
curvature,  spacetime structure itself is undetermined until the
object's motion is determined. Instant by instant,  spacetime
structure changes from indeterminate to definite. Thus a definite
spacetime structure comes into being as time evolves.  Spacetime is
unknown and unpredictable before it is determined. The Evolving
Block Universe model of spacetime represents this situation: time
progresses, events take place, and history is shaped. Second by
second, one specific evolutionary history out of all  possibilities
is chosen, takes place, and becomes cast in stone. The basic EBU
idea was proposed many years ago by Broad \cite{Bro23}, but has not
caught on in the physics community. The new element in the more
recent presentation \cite{Ell06} is the random element introduced
through the irreducible uncertainty of quantum events. This ensures
that there is no way the future spacetime can be predicted from the
past: what will actually happen is not determined until it happens.

The EBU can be represented through a growing spacetime diagram, in
which the passing of time marks the change from  indefinite (not yet
existing) to definite (having come into being), and in which  the
present marks the instant at which we can act and change reality
(see Figure 2). Even the nature of future spacetime is taken to be
uncertain until it is determined at the ``now," along with the
physical events that occur in it.  Unlike special relativity, which
operates in matter-free space, the EBU assumes that matter exists
and causes spacetime curvature. Depending on the distribution of
matter and energy, particular spacetime surfaces and timelike
worldlines will be geometrically and physically preferred. The
solutions embody broken symmetries of the Einstein field equations:
the solutions have less symmetry than the equations of the theory.

One is led to suggest \cite{Ell06} that the transition from present
to past does not take place on specific spacelike surfaces; rather
it takes place pointwise at each spacetime event, with dynamical
processes taking place along timelike world lines (associated with
possible particle trajectories), rather than on spacelike surfaces
(implicit in usual presentations of the initial value problem in
General Relativity).

%--------------------------------------------------------------------------------------------------------

\begin{figure}[htb] \vbox{\hfil\scalebox{.5}
{\includegraphics{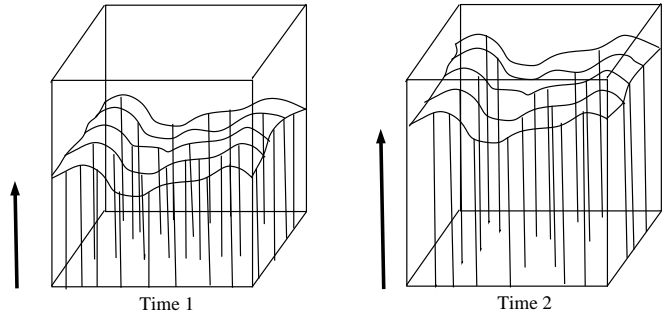}}\hfil} {\caption{\footnotesize{An evolving curved space-time picture that takes
macro-phenomena seriously. Time evolves along each world line,
extending the determinate spacetime as it does so. The particular
surfaces have no fundamental meaning and are there  for convenience
only (one requires some coordinates to describe what is happening).
You cannot locally predict uniquely to either the future or the past
from data on any `time' surface (even though the past is already
determined). This is true both for physics, and (consequently) for
the spacetime itself: the developing nature of spacetime is determined
by the evolution (to the future) of the matter in it.
 \label{fig2}}}}
\end{figure}

%--------------------------------------------------------------------------------------------------------\\

The proposed view is thus that spacetime is continually extending to
the future as events develop along each world line in a way
determined by the complex of causal interactions; these shape the
future, including the very structure of spacetime itself, in a
locally determined (pointwise) way.  The Evolving Block Universe
continues evolving along every world line until it reaches its final
state, resulting in an unchanging Final Block Universe. One might
say that then time has changed into eternity. The future is
uncertain and indeterminate until local determinations have taken
place at the spacetime event ``here and now," designating the
present on a world line at a specific instant; thereafter this event
is in the past, having become fixed and immutable, with a new event
on the world line designating the present. %There is no unique way to
%describe this process relatively for different observers;
Analysis of the evolution is conveniently based on preferred (matter
related) world lines rather than time surfaces. However in order to
describe it overall, it will be convenient to choose specific time
surfaces for the analysis, but these are a choice of convenience
rather than necessity.  While the general coordinate invariance
invoked in general relativity theory proclaims there are no
preferred such surfaces, in any particular solution this is not the
case - there will for example usually be preferred spacelike
surfaces in any particular cosmological solution.

\emph{In summary}: The past has taken place and is fixed, and so the
nature of its existence is quite different than that of the
indeterminate future. Uncertainty exists as regards both the future
and the past, but its nature is quite different in these two cases.
The future is uncertain because it is not yet determined: it does
not yet exist in a physical sense. Thus this uncertainty has an
ontological character. The past however is fixed and unchanging,
because it has already happened, and the time when it happened
cannot be revisited; but our knowledge about it is incomplete, and
can change with time. Thus this uncertainty is epistemological in
nature.

%-----------------------------------------------------------------------------
\section{Taking quantum physics seriously: emergence of certainty}
\setcounter{equation}{0}\label{sec3}
%-----------------------------------------------------------------------------

Quantum physics is, in crucial respects, very different from
classical physics and we have to now take this into account. Quantum
phenomena underlie classical phenomena, but often with completely
counter-intuitive features. At the same time classical behaviour as
we know it reliably emerges at everyday scales.

%-----------------------------------------------------------------------------
\subsection{Quantum features}
\setcounter{equation}{0}\label{sec3.1}
%-----------------------------------------------------------------------------
In the quantum world
\cite{AhaRoh05,GreZaj06,LonBau83,Fey85,Pen89,Mor90,Leg91,Rae94,Ish97,BrePut06}
we must reconcile ourselves to that fact that:

$\bu$  Only probabilities can be predicted, not specific outcomes;

$\bu$  These probabilities are determined by a complex wave function
or state vector, which evolves in a unitary fashion (a
time-reversible linear process);

$\bu$ The probabilities of measurement outcomes are given by the
square of this state vector (this is a key non-linear feature of the
theory);

$\bu$ Superposition of states is allowed, indeed required, this
process enabling interference and  quantum tunneling to take place,
and generically leading to quantum entanglement emerging;

$\bu$  In an entangled state,\footnote{Two states are said to be
``entangled" when they cannot be viewed as being independent of each
other in any way, technically when their wavefunctions cannot be
written as a simple product state.} state physical entities do not
have separate individual existence; rather it is a collective state
described by macroscopic variables;

$\bu$  The potentialities of the quantum state are converted into
the actualities of the classical state by a quantum measurement
process that is not adequately described by current quantum theory;
its existence and nature is a postulate over and above that of the
current theory;

This process generically has a two-part nature:

\begin{quotation}
$\bu$ First is a process of environmental decoherence describable
via unitary evolution, which effectively disentangles states and
leads to the emergence of individual properties (decoherence
converts entangled states to mixtures).  The resulting state,
however, while successfully predicting the probabilistic outcomes of
statistical series of events, fails to predict the unique outcome of
any specific event: it still represents a set of as yet unfulfilled
potentialities;

$\bu$ These potentialities are converted to a specific outcome by
the second part of the measurement process:``collapse of the wave
function" or ``state vector reduction."   This stage is where the
central feature of quantum uncertainty manifests itself (this
uncertainly never occurs in processes described only by the unitary
evolution equations); thus this process can be thought of as
actualizing potentialities;

$\bu$ This second process is time-irreversible and causes
information loss, and so is not describable by any unitary
evolution;

$\bu$  The two-part process is the generic case; however in specific
cases, unique outcomes can occur where one or other of these
measurement processes is not needed (i.e. classical outcomes can
occur via unitary evolution).  For example decoherence is not needed
if the initial state is not entangled, and ``wave function collapse"
will not take place if  the state vector is already in an eigenstate
of the chosen operator;

$\bu$ The measurement process resolves the dualities of quantum
theory through choosing specific determinate outcomes, thus leading
to classical states;

$\bu$ It does so in a way that effectively can influence events that
are apparently in the past, as is made manifest in delayed choice
experiments (see \S \ref{sec4});

$\bu$ The process of state preparation shares many of the properties
of the process of measurement.
\end{quotation}

The upshot of this all is that weird quantum properties characterize
matter before wave reduction take place; after it has taken place,
the potential in its prior state is converted into definite
outcomes.  But the evolutionary equations of quantum theory are
unitary, and hence entirely deterministic. Therefore the measurement
process by which uncertainty is transmuted to certainty goes
entirely undescribed by the theory itself.  It is an extra feature
outside quantum mechanics, but is essential for testing and
interpreting the theory.  Because the measurement process is crucial
for the CBU picture, we devote the next section to outlining it.

%-----------------------------------------------------------------------------
\subsection{Determinate to indeterminate}
\setcounter{equation}{0}\label{sec3.2}
%-----------------------------------------------------------------------------

The above considerations underscore the fact that we can't  uniquely
predict the future because of foundational quantum uncertainty
relations (see e.g. \cite{Fey85,Pen89,Ish97}). We cannot predict
precisely when a nucleus will decay or what the velocity of the
resultant particles will be, nor can we predict precisely where a
photon or electron in a double-slit experiment will end up on the
screen.  This unpredictability is not a result of a lack of
information: it is the very nature of the underlying physics. More
formally, consider the wave function or state vector $\ket\ps(x)$.
The basic expansion postulate or quantum mechanics is that before a
measurement is made, $\ket\ps$ can be written as a linear
combination of eigenstates
 \be
 \ket{\ps_1} = \sum_n c_n\ket{u_n(x)}, \label{wave}
 \ee
where $u_n$ is an eigenstate of some observable $\hat A$ (see e.g.
\cite{Ish97}: 5-7).  Immediately after a measurement is made at a
time $t=t^*$, however, the wavefunction is found to be in one of the
eigenstates:
 \be
 \ket{\ps_2} = c_Nu_N(x) \label{collapse}
 \ee
for some specific index $N$.  The data for $t< t^*$ do not
determine $N$; they merely determine a probability for the outcome
$N$ through the fundamental equation
 \be
p_N = c_N^2.  \label{prob}
 \ee
One can think of this as due to the probabilistic time-irreversible
reduction of the wave function\\
\be
 \begin{array}{lll}
 \ket{\ps_1} = \sum_n c_n\ket{u_n(x)} \hspace{1.75 cm}
 \longrightarrow \hspace{1.5 cm}\ket{\ps_2} = c_Nu_N(x)\nn\\
\hspace{.5 cm} Future: \hspace{2.75 cm} Present:   \hspace{1.5 cm} Past: \nn\\
 Indeterminate  \hspace{2 cm} Transition  \hspace{1 cm} Determinate \label{trans}
\end{array}
\ee \\
(\cite{Pen89}: 260-263). This is the event where the uncertainties
of quantum theory become manifest (up to this time the evolution is
determinate and time reversible). Invoking a many-worlds description
(see e.g. \cite{Ish97}) will not help: in the actually experienced
universe in which we make the measurement, $N$ is unpredictable
(this proposal does not clarify in which branch of the wave function
any specific observer will end up, so the experimental outcome
(\ref{collapse}) is unaltered by this hypothesis).

A hidden variable theory may help, but here we will deal with
standard quantum theory as determined by experiments. Thus the
initial state (\ref{wave}) does not uniquely determine the final
state (\ref{collapse}); and this is not due to lack of data, it is
due to the foundational nature of quantum interactions. You can
predict the statistics of what is likely to happen but not the
unique actual physical outcome, which unfolds in an unpredictable
way as time progresses; you can only find out what this outcome is
after it has happened. Furthermore, in general the time $t^*$  is
also not predictable from the initial data: you don't know when  the
transition from (\ref{wave}) to (\ref{collapse}) will take place,
because you can't predict when a specific excited atom will emit a
photon, or when a radioactive particle will decay.

We also can't retrodict to the past at the quantum level, because
once the wave function has collapsed into an eigenstate it is
impossible to determine from that eigenstate the configuration
before the measurement. The fact that such events happen at the
quantum level does not prevent them from having macro-level effects.
Many systems can amplify quantum effects to macro levels, including
photomultipliers (whose output can be used in computers or
electronic control systems). Quantum fluctuations also change the
genetic inheritance of animals, and so influence the course of
evolutionary history on Earth. This is effectively what occurred
when cosmic rays-- -whose emission processes are subject to quantum
uncertainty---caused genetic damage in the distant past
\cite{Ell06}. Furthermore, similar processes not only affect events
in spacetime, but influence spacetime itself.  For example, shortly
after the big bang quantum fluctuations were amplified to
astronomical scales by the universe's expansion, becoming seeds for
galaxy formation (see \cite{Ell06}).

As regards the CBU, the key feature is that not all transitions in
quantum physics can be characterized by time-reversible Hamiltonian
dynamics. This fundamental problem does not become evident if one is
content to consider only statistical predictions for ensembles of
identical microparticles: in that case environmental decoherence
\cite{Kie02} results in diagonalization of the density matrix,
giving the same statistical predictions for quantum and classical
systems. But in the case of distinguishable particles where we want
to determine the specific outcome for a single system, statistical
predictions are inadequate and the collapse of the wave function to
a specific eigenstate becomes germane; the collapse, as is well
known, cannot be described by Hamiltonian dynamics.

%-----------------------------------------------------------------------------
\section{Delayed choice experiments: Affecting the past}
\setcounter{equation}{0}\label{sec4}
%-----------------------------------------------------------------------------

Secondly, and more controversially, the collapse itself takes place
in a way that, at least insofar as conventional language goes,
appears to allow a degree of influence from the present to the past,
reflecting the time-symmetric nature of the underlying physics. This
is such an important feature of the Crystallizing Block Universe
that we devote this section to the experimental evidence for such a
claim and the next to theoretical attempts to come to grips with
this phenomenon.

%-----------------------------------------------------------------------------
\subsection{Wheeler's delayed choice experiments}
\setcounter{equation}{0}\label{subsec4.1}
%-----------------------------------------------------------------------------

The double slit experiment and its derivatives, particularly
Wheeler's delayed-choice experiment \cite{Whe78}, show that we are
apparently able to influence the past to some degree, or at least
that the conventional notion of ``the past'' is ill defined.   The
classic form of Wheeler's gedanken-experiment involves the usual
double-slit configuration, in which a photon passes through the
slits to fall on a detector or screen at some later time.  As in the
standard double-slit experiment, one can choose to measure the wave
or particle property of the photon: measuring the photon's momentum
(particle property) allows determination of the slit through which
it passed but wipes out any information about its position (the
interference pattern on the screen, a wave property).   Measuring
the interference pattern, on the other hand, destroys any
information about the photon's momentum.

Wheeler's version of the experiment \cite{Whe78} added the striking
feature that if the distance between the slits and the detector is
sufficiently large, one can choose to measure the particle or wave
property {\it after} the photon has passed the slits. How can we determine
the photon's properties, after they should have already been
decided? As counterintuitive as the situation may seem, over the
past two decades laboratory confirmation of Wheeler's delayed-choice
experiment has been achieved, in particular by the 2006 experiment
of Jacques et al. \cite{Jacetal07}, who attained levels of one
photon in the apparatus at a given time, ruling out possible
confusion in counting properties of different photons.

%-----------------------------------------------------------------------------
\subsection{Quantum erasers}
\setcounter{equation}{0}\label{subsec4.2}
%-----------------------------------------------------------------------------

An equally, if not more bizarre, variant on the two-slit experiment
proposed in 1982 by Scully and Dr\"uhl \cite{ScuDru82} has also been
realized in the laboratory. The ``quantum eraser'' allows one to
measure ``simultaneously" both the slit through which a photon has
passed and the interference pattern.  To be sure, the position
information (interference pattern) disappears when momentum (``which
slit'') information is obtained, but it reappears when the
``which-slit'' information is erased.  The position information has
not been destroyed by the momentum measurement, in contradiction to
conclusions traditionally drawn from the Heisenberg uncertainty
principle \cite{AhaZub05,Cra00,Rho00}.  Moreover, the quantum eraser
has also been performed in a delayed-choice version
\cite{Kimetal00}, in which the interference pattern can be erased or
recovered even ``after'' the photon has been detected.

%-----------------------------------------------------------------------------
\subsection{Weak measurements}
\setcounter{equation}{0}\label{subsec4.3}
%-----------------------------------------------------------------------------

A perhaps more well-known variety of experiment that throws doubt on
the conventional interpretation of the past are the so-called ``weak
measurements'' originally proposed by Aharonov, Albert and Vaidman
\cite{AhaAlbVai86}. Invariably, students are taught that any
measurement of a quantum system collapses the wave function into an
eigenstate. Weak measurements, as their name implies, disturb the
system so little that no collapse in fact takes place.  Furthermore,
one is accustomed to think of the outcome of an experiment being
determined by initial conditions.  In quantum mechanics initial
conditions are insufficient to determine the outcome of an
experiment, due to the probabilistic nature of the theory; one needs
both initial and final conditions to gain full knowledge.   In
devising a weak measurement, one prepares the system in a certain
state (makes a ``preselection''), later makes a final strong
measurement (``postselection'').  Both these initial and final
boundary conditions determine the outcome of an intermediate
(``weak'') measurement.   For example, if an ensemble of identical
subsystems is preselected at an initial time $t_0$ such that $S_z =
\up$  and postselected at a later time $t_2$ such that $S_x = \up$,
then any measurement on an individual subsystem at an intermediate
time $t_1$ must reveal $S_z = \up$ (it was prepared to be so) as
well as $S_x = \up$ (otherwise the final measurement at $t_2$ would
not get this result). For an arbitrary angle $\ph$ in the x-z plane,
however, $S_\ph = S_z\sin\ph + S_x\cos\ph$.  But $S_z = S_x = 1/2$,
and so a measurement at $\ph = \p/4$ will yield $S_\ph = \sqrt 2/2$,
an apparently impossible result.

For textbook ``strong" measurements, this is indeed impossible
because $S_z$ and $S_x$ are non-commuting variables and hence a
measurement on $S_x$ should ``disturb'' a measurement on the initial
state, which is not an eigenstate of $S_x$.  If, however, the
initial state is indeed an ensemble consisting of many
noninteracting replicas of the same subsystem (a product state),
then a measurement on a single $S_x$ will be ignored by the vast
majority of subsystems and the overall wavefunction will remain
arbitrarily close to its initial state. The penalty one pays for
registering both components of the spin is that the pointer of the
device moves more than it should---the measurement looks like an
error. Although one might balk at the notion of a final measurement
influencing an intermediate measurement, the fact is, like the
delayed-choice and eraser experiments, weak measurements have been
experimentally observed (e.g. \cite{HosKwi08}) and, indeed, have
recently provided a high degree of amplification in optical systems
\cite{Dixetal09}.

%-----------------------------------------------------------------------------
\section{Affecting the past: Quantum proposals}
\setcounter{equation}{0}\label{sec5}
%-----------------------------------------------------------------------------

The issue of delayed choice arises in regard to measurements that
are made, and of course we don't yet have a fully satisfactory
theory of quantum measurement: how the unitary evolution of the
state vector changes to a single eigenstate when an observation is made.
  By contrast the unitary evolution between state
measurements is well understood, as are the eigenvalues associated
with particular outcomes, e.g. energy levels and scattering angles,
and these are the main topic of most quantum theory textbooks. The
issue then becomes that unitary time evolution is  time
symmetric: influences from the future should be as effective as
influences from the past.

Some experiments, mentioned above, suggest that there are indeed
real physical effects related to the time-reversed solutions: that
in fact we can to some degree act on the past. Various theoretical
efforts try to show how the measurement process could lead to such
partially time symmetric outcomes. We here briefly mention two main
classes of these theories; in a separate paper will analyze more
closely their technical details and relationship to one another.

As a preliminary remark, we emphasize that any complete attempt to
relate physical theory to the macro world must ultimately deal with
\emph{individual systems and measurements} as well as ensembles,
because specific things happen to individual entities in the real
world. Theories that deal with statistical predictions for ensembles
of objects are very helpful in many contexts, but they simply do not
succeed in giving unique predictions for specific objects; and a
complete theory of the world should be able to do that. That is why
an attempt to relate the theory to experiments, involving state
vector reduction to a specific eigenstate, is essential.

%-----------------------------------------------------------------------------
\subsection{Transactional interpretations of quantum mechanics}
\setcounter{equation}{0}\label{subsec5.2}
%-----------------------------------------------------------------------------
%-----------------------------------------------------------------------------
%\subsubsection{Wheeler and Feynman: Advanced and retarded solutions}
%-----------------------------------------------------------------------------

Many of the attempts to come to grips with the time-symmetric nature
of quantum mechanics find their roots in the Wheeler-Feynman (WF)
absorber theory of electromagnetic radiation
\cite{WheFey45,WheFey49}, which respects the time-symmetry inherent
in Maxwell's equations. Developing earlier work by Tetrode and
others, Wheeler and Feynman proposed a model where the field
generated by an accelerated charge should consist of one-half the
advanced plus one-half the retarded Li\'enard-Wiechert solutions of
Maxwell equations.  Thus, at $t=0$ the accelerated source charge
emits a field that is $1/2(advanced + retarded)$.  The retarded part
reaches an absorber located at a distance $r$ after a time $r/c$.
The absorber generates a field that is  $1/2(retarded -  advanced)$,
which converges on the source at $t=0$.  In other words, the
advanced part of the wave generated by the absorber is $180^\circ$
out of phase with the advanced part generated by the source.
Consequently, the advanced parts cancel and we are left with the
usual retarded solution.

The WF theory is totally time-symmetric in that an advanced wave may
be reinterpreted as a retarded wave by reversing the sign on $t$ and
vice-versa.  WF concluded that the irreversibility of the emission
process must be ``a phenomenon of statistical mechanics connected
with the asymmetry of the initial conditions with respect to time."
From their considerations, in particular of ``pre-acceleration,"
Wheeler and Feynman conclude that ``the past and future of all
particles are tied together by a maze of interconnections"
\cite{WheFey45}. Nevertheless, Feynman later conceded \cite{Fey49}
that the theory failed to give an adequate account of the
self-energy of the electron, and it has been largely abandoned.

%-----------------------------------------------------------------------------
%\subsubsection{Cramer: Transactional interpretation of quantum mechanics}
%-----------------------------------------------------------------------------
Although the WF theory was a theory of electromagnetism, Cramer
\cite{Cra86,Cra88} has taken over its essentials for his
``Transactional interpretation of quantum mechanics". He adopts
$\ps^*(t)$, the solution to the conjugate equation, as the advanced
wavefunction. Analogously to the  WF picture, a quantum event is
associated with a completed transaction.  The retarded state vector
extends an ``offer wave," with amplitude $\ps(t)$.  The absorber
replies with an advanced ``confirmation wave," $\ps^*(t)$, giving an
amplitude back at the source (assuming no attenuation) of
$P=\ps^*\ps$, which would be interpreted as the probability of the
event.  If there are many absorbers, the sum of all such
offer-confirmation echoes is $\int \ps^*\ps d\t$, the overlap
integral.

The transactional interpretation can account for many quantum
``paradoxes" in a fairly straightforward fashion. In the case of
Wheeler's delayed-choice experiment, assume that if one wants to
measure the wave properties of the photon, a photographic emulsion
can be ``instantly" put in place; to measure the momentum
properties,  the emulsion can be instantly lowered to reveal highly
collimated CCD detectors, which record the ``which-slit"
information.  For a position measurement in the transactional
interpretation, a retarded offer wave travels through both slits to
be absorbed at the emulsion.  An advanced confirmation wave retraces
the path through both slits to the source.  Only ``after" the
confirmation wave reaches the source, the transaction is completed
and the photon is said to have passed through both slits.  In the
complementary situation, the offer wave as before passes through
both slits, but now the emulsion is lowered and the photon is
recorded by one of the CCD detectors.  The confirmation wave passes
backwards through only one slit and the completed transaction is
that characteristic of a single-slit event.

While the transactional interpretation is a bold development of
the WF proposal, there are some difficulties with its interpretation; we
deal with these in our technical companion paper.

%-----------------------------------------------------------------------------
\subsection{Two-time interpretations of quantum mechanics}
\setcounter{equation}{0}\label{subsec5.1}
%-----------------------------------------------------------------------------
These interpretations are based in the idea, mentioned above, that
one often deals with both pre- and post-selected ensembles,
associated with state preparation and state measurement. For
example, scattering experiments involve  an  initial
beam, prepared to contain particles with a specific energy and momentum,
 and after the scattering event each detector measures the outgoing particles in another
specific direction and energy range. Thus both preparation and
measurement project into pre-determined eigenspaces of Hilbert space
(note that the relevant operations do not necessary relate to a
complete description of the state: they may leave some variables
undetermined). The duality between pre- and post-selection reflects
the time symmetry of the underlying theory.

%-----------------------------------------------------------------------------
%\subsubsection{Aharonov and Gruss: Two-time interpretation of quantum mechanics}
%-----------------------------------------------------------------------------

In 1964, Aharonov, Bergmann and Lebowitz (`ABL') \cite{ABL}
investigated the assertion that the ``collapse of the wavefunction"
introduces an arrow of time at a fundamental level in quantum
mechanics.  They found that, to the contrary, the arrow of time is
introduced by the way statistical ensembles are constructed, and if
time-symmetric boundary conditions are introduced, the resulting
probability distributions are time symmetric as well. The ABL
formula \cite{AhaRoh05,ABL} gives the probability that an
intermediate measurement (between an initial and a final time)
results in an eigenstate $a_k$ assuming that both initial and final
states were specified:

\be prob(a_k|\ps_i,\ps_f) =
\frac{|\media{\ps_f|a_k}|^2|\media{a_k|\ps_i}|^2}{\sum_j|\media{\ps_f|a_j}|^2|\media{a_j|\ps_i}|^2},
\label{ABL} \ee

\noindent where $\ps_i$ is the initial state and $\ps_f$ is the
final state. Eq. (\ref{ABL}) is the product of two conditional
probabilities, the first giving the probability of getting $a_k$
given $\ps_f$ and the second giving the probability of getting $a_k$
given $\ps_i$.  This formula shows that, to be sure, if final as
well as initial conditions are specified, then the probability
distribution is time symmetric. As already discussed in regard to
weak measurements in \S \ref{sec4},  which were an outgrowth of the
ABL formula, one  specifies final boundary conditions in the same
way as initial conditions: by selecting some members an ensemble on
the basis of a specified property, eg. all electrons that are spin
up at some final time. In the language of weak measurements one
``preselects" initial states and ``postselects" final states. In
some sense this is a restatement of Gibbs' famous objection to the
usual statistical mechanical argument for the increase of entropy.
As Gibbs pointed out, given any low entropy state, entropy increases
in both time directions; we, however, generally cut off the past; we
predict rather than retrodict.

The ABL paper has been the subject of some controversy; we point out
here only that Shimony \cite{Shi05} has vindicated the proposal by
a rigorous derivation using Bayes theorem together with standard
quantum mechanical predictions regarding ensembles that are only
pre-selected.

 In 2005 Aharonov and Gruss (AG, \cite{AhaGru05}; see also Section 18.3 of \cite{AhaRoh05})
announced a two-time interpretation of quantum mechanics,
developing from the work of ABL.  Due to the first-order nature of
the Schr\"odinger equation, one only has a single initial (or
final) condition at one's disposal and so, if we impose both, we
must have one state vector evolving forward in time and another
evolving backward in time. They propose a History state vector
$\ket{\Psi_{HIS}(t)}$ determined by the initial state
$\ket{\Psi_{HIS}(t_0)}$ of the system and a Destiny state vector
$\bra{\Psi_{DES}(t)}$ determined by its final state
$\bra{\Psi_{DES}(t_f)}$. From the History and Destiny vectors they
form the ``two-state'' density operator

\be \rho(t) = \frac{\ket{\Psi_{HIS}(t)}\bra{\Psi_{DES}(t)}}
{\bra{\Psi_{DES}(t)}\ket{\Psi_{HIS}(t)}} \label{twostate}\ee

\noindent AG furthermore postulate that the system can be completely described
by this operator, which evolves as

\be \hat\r(t_2) = U(t_2,t_1)\r(t_1)U(t_1,t_2) \ee

\noindent where $U(t_2,t_1)$ is the standard unitary evolution
operator $U=e^{-iH(t_2-t_1)}$.

The AG paper is in effect a restatement and development of a much
earlier proposal by Davidon \cite{Dav76}; the relation between
the two works will be discussed in our further paper.
AG use the formalism to examine a spin-1/2 system from the two-time perspective and conclude that
the description is entirely deterministic and local.  They also show that
two-time decoherence is completely time-symmetric, and go on to  examine
the implications for the two-time scenario when the final states are
randomly distributed, leading to the possibility the  weak
measurements  discussed in  \S \ref{sec4}.

AG claim that the two-time formalism solves the quantum measurement
problem: from their abstract, ``the quantum superposition is, in
effect, dynamically reduced to a single classical state via a
`two-time decoherence' process'.'' It is true that by imposing a
final condition (say ``up" for the apparatus in the destiny vector)
one finds that the ``two-state" operator (\ref{twostate}), and hence
the state of the system, must be measured to be ``up" without any
nonunitary evolution having taken place. However in order to recover
the usual quantum probabilities during measurements on an ensemble,
AG must assume a random distribution of final conditions. The
assumed distribution of final conditions now determines the outcome
of the measurements, and this is put in by hand.  Thus, the AG
proposal does not appear to adequately resolve the measurement
problem, although it must be reemphasized that, regardless of
interpretation, weak measurements have been observed.

 More recently Davies \cite{Dav08} has used the
two-time formalism to show that the textbook exponential decay law
is in fact a restricted case of a more general quantum decay law
that depends on both the time of post-selection and the final state,
and has proposed that post-selection from the final state of the
universe can be used to explain important features of the universe
\cite{Dav06}.

%-----------------------------------------------------------------------------
\subsection{Overall}
\setcounter{equation}{0}\label{subsec5.3}
%-----------------------------------------------------------------------------
From the previous discussions, it is apparent that a number of serious proposals
have been put forth that attempt to relate the time-symmetric
unitary evolution of quantum theory to the possibility
of the future influencing the past.  Furthermore some of these
theoretical explorations have led to new and notable experimental
results. These attempts, however, do not express the passage of time
as the EBU does.   Many of them in fact assume that the
future already exists (if the future did not already exist, it would
of course not be able to influence the present or past). Our claim,
by contrast, is that the future does not yet exist; at present the future
is merely a set of possibilities. In the spirit of the EBU, any
spacetime picture must be adopted to take this into account, as
well as take into  account the demands of quantum physics. Once one
has this revised spacetime picture in mind, it should be possible to
revisit all these theories and adapt them to a more realistic
spacetime description.  In the following sections we undertake to outline a
spacetime description that unifies the backward-causality demonstrated in various
quantum experiments with a realistic view of spacetime.

%-----------------------------------------------------------------------------
\section{The Crystallizing Block Universe}
\setcounter{equation}{0}\label{sec6}
%-----------------------------------------------------------------------------
The challenge is to modify the EBU picture of \S \ref{sec2}, to give
a spacetime picture in which the paradoxical foundational features
of quantum theory are taken seriously.   In order to do so, we
contemplate a crystallizing nature for the emergence of spacetime:
not all features become fixed at the same time, and post-selection
of previous events is possible. Potentiality changes to actuality at
each quantum measurement process, but some potentialities may remain
undecided even as others have transmuted to definiteness. Thus we
consider that on a given world line ``now"  is the moment when those
aspects of reality become fixed. On neighboring world lines the
transition may occur at a later time and until the the transition
takes place such aspects remain indefinite.  The idea of `post
selection' then applies \emph{at the present time}: we can
\emph{now} post-select states relating to interactions that have
taken place in the past (as is made clear in the examples discussed
by ABL).

The EBU evolves as discussed above, but does not settle all issues
at once: some ``events" that are left behind in the ``past" remain
uncertain, to become fixed only at a later time. An everyday
analogy would be one of a crystallizing molten mixture: as the
lattice of material becomes hardened and defines a fixed
structure, some molten bits remain in the interstices to become
fixed only later. The large scale structure crystallize out first
and the inner details are filled in later; quantum uncertainty
applies more to small objects than larger ones. The detailed past
and future are separated by this crystallization surface (``CS"),
rather than any macro approximation we make to it through a
coarse-grained ``surface of constant time."

The relation of quantum physics to the real universe is completely
different to the past and the future of this surface: if we imagine
a part of spacetime extending to the future of the CS, as a ghostly
forerunner of the CBU spacetime domain, this region is subject to
quantum uncertainty as expressed in the foundational probability
equations (\ref{prob}) of quantum theory, and hence to all the
associated feature of quantum duality. The part of spacetime to the
past is not subject to any quantum uncertainty: all such uncertainty
has been resolved, so the fundamental equation (\ref{prob}) is no
longer applicable in that part of spacetime. It is replaced by the
equation
 \be
p_n = 1\, (n = N),\,\, p_n = 0\, (n\neq N).  \label{prob1}
 \ee
Thus quite different probabilistic equations hold in the past and
the future of the CS (if we imagine spacetime extended to the
future). This is why the ontological status of each domain is
completely different.

%----------------------------------------------------------------

\begin{figure}[htb] \vbox{\hfil\scalebox{.5}
{\includegraphics{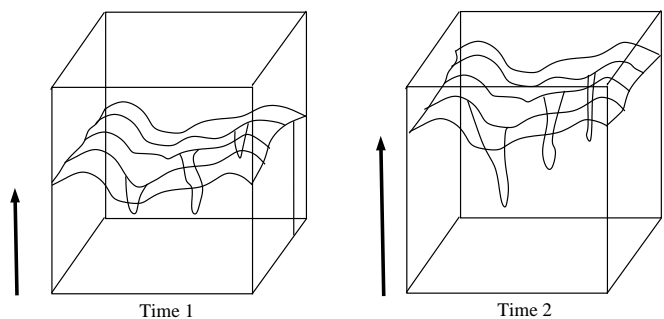}}\hfil} {\caption{\footnotesize{An evolving curved space-time picture that
takes micro-phenomena seriously. Like the EBU (Figure 2), but here
small pockets of potentiality remain unresolved till later times:
delayed choice experiments can influence these `past' events. The
resolution of potential to definite outcomes takes place in a way
like a crystallizing mixture, with pockets of residue remaining
behind for a time after the main resolution front has passed.
 \label{fig3}}}}
\end{figure}

%-----------------------------------------------------------------------------
%\subsection{Levels of detail}
%\setcounter{equation}{0}\label{subsec6.1}
%-----------------------------------------------------------------------------
The structure of spacetime is scale-dependent: when averaged on a
large enough scale, this detailed micro-structure will be invisible,
so on a coarse-grained view, we regain the classical EBU as
described above. The description obtained thus depends on the
%(coarse-grained)
level of detail one represents, with an averaged surface of
concretization for the macro level (the macroscopic time ``now'')
differing in detail from the microscopic surface CS where quantum
uncertainty gives way to definiteness in a point by point way.

%-----------------------------------------------------------------------------
\subsection{Experiments and Theory underlying the CBU}
\setcounter{equation}{0}\label{subsec6.2}
%-----------------------------------------------------------------------------
The experiments underlying the CBU have already been described
above, in \S \ref{sec4}. They seem to adequately validate the
picture being put forward here, indeed they demand such a picture,
whether we can provide an adequate theoretical explanation for these
kinds of events or not.

Is there a viable theory underlying the CBU concept? A series of
 proposals have been described in \S \ref{sec5} above.
In brief: firstly, one can suggest that entanglement takes place in
space and time, rather than just in space; secondly, decoherence can
act apparently into the past as well as at a distance; and thirdly
either a two-time or a transactional version of quantum theory
promises to give an adequate theoretical underpinning for the
experiments.

However there is one major factor to now be taken into account:
those theories did not envisage a CBU context, and it is interesting
to see how they must change when the CBU is introduced.  Many of the
proposals talk about final conditions for the universe; but we can
only assign final conditions at the present time. Thus we envisage
the possibility that either a transactional or a two-time formalism
could underlie our proposal in a satisfactory way, but with `` final
conditions'' imposed at the present time, instead of at the not-yet
existing end of the universe. That suggestion is indeed largely
satisfied already by the mathematical analyzes underlying the two
sets of proposals, where there is in fact no specific mathematical
feature characterizing the envisaged  ``final state" as necessarily
existing ``at the end of time."

Thus we suggest those analyzes can be adopted more or less as is,
with the ``final state" being taken as the present time $t_0$. Thus,

$\bu$ It may possibly proceed as a variant of that described by
Aharonov and Gruss \cite{AhaGru05}, but with the later time in that
formalism everywhere changed to the present day rather than the
final state of the Universe, thus making their interpretation
compatible both with the EBU idea and with realistic experimental
testing.

$\bu$ It may alternatively already be largely contained in the
transactional viewpoint put by Cramer \cite{Cra86,Cra88}, but now
interpreted in the CBU context.

These proposals will be explored further in our follow-up paper.

%-----------------------------------------------------------------------------
\subsection{Measurement and the CBU}
\setcounter{equation}{0}\label{subsec6.4}
%-----------------------------------------------------------------------------
The measurement interaction may perhaps be regarded as an
interaction between scales. The measurement process itself is the
crucial transition from indefinite to definite. On the face of it,
this has the nature of top-down action from the macro level to the
micro level \cite{Ell08},  as is explicitly stated in the Copenhagen
interpretation of quantum theory, and is implicit in most quantum
physics writings (e.g. \cite{Fey85}), where a macroscopic detector
is assumed to exist and give definite results. Decoherence
\cite{Kie02} also has a top-down nature, expressed in the idea of
environmental selection (`einselection') \cite{Zur04}.

Thus the CBU may provide a sound context for expressing top-down
causation from the macro to the micro level inherent in the
measurement process, effective at the present time but with traces
remaining undetermined until a later time. How this may be
realised in detail  needs to be pursued.

%-----------------------------------------------------------------------------
\subsection{The arrow of time}
\setcounter{equation}{0}\label{subsec6.5}
%-----------------------------------------------------------------------------

As regards the arrow of time problem \cite{Dav74,Zeh92}, if the CBU
view is correct, the Wheeler-Feynman prescription for introducing
the arrow of time by integration over the far future
\cite{WheFey45}, and associated views comparing the far future with
the distant past \cite{Pen89,EllSci72,Ell02}, are  invalid
approaches to solving the arrow of time problem, for it is not
possible to do integrations over future time domains if they do not
yet exist. Indeed the use of half-advanced and half-retarded Feynman
propagators in quantum field theory then becomes a calculational
tool representing a local symmetry of the underlying physics that
does not reflect the nature of emergent physical reality, in which
that symmetry is broken.

The arrow of time problem needs to be revisited in this CBU context.
The key point is that the arrow of time arises simply because
\emph{the future does not yet exist}. One can be influenced at the
present time from many causes lying in our past, as they have
already taken place and their influence can thereafter be felt. One
cannot be influenced by causes coming from the future, for they have
not yet come into being.  The history of the universe has brought
the past into being, which is steadily extending to the future, and
the future is just a set of unresolved potentialities at present.
One cannot integrate over future events to determine their influence
on the present not only because they do not yet exist, but because
they are not even determined at present (as explained above in
Section \S \ref{sec2}).

The direction of the arrow of time is thus determined in a
contingent way in the CBU context. Collapse of the quantum wave
function is a prime candidate for a location of a physical solution
to the arrow-of-time problem, and manifests itself as a form of
time-asymmetric top-down action from the universe as a whole to
local systems (cf. \cite{Pen89}). This takes place within the
generic context of commonly occurring top-down action in the
hierarchy of causality \cite{Ell08}.

%-----------------------------------------------------------------------------
\section{Overall: A more realistic view}
\setcounter{equation}{0}\label{sec7}
%-----------------------------------------------------------------------------
The nature of the future is completely different from the nature of
the past.  When quantum effects are significant, the future
manifests all the signs of quantum weirdness, including duality,
uncertainty, and entanglement. With the passage of time, after the
time-irreversible process of state-vector reduction has taken place,
the past emerges, with the previous quantum uncertainty replaced by
the classical certainty of definite particle identities and states.
The present time is where this transition largely takes place. But
the process does not take place uniformly or reversibly:  evidence
from delayed choice experiments shows that some isolated patches of
quantum indeterminacy remain, and their transition from probability
to certainty only takes place later. Thus, when quantum effects are
significant,  the Evolving Block Universe (``EBU") of classical
physics cedes way to the Crystallizing Block Universe (``CBU"). On
large enough scales that quantum effects are not significant, the
two models become indistinguishable.

\emph{\textbf{Acknowledgements:}} We thank the NRF (South Africa)
for support that made our collaboration on this project possible.\\

%-----------------------------

-------------------------------------------------------------------------------------
\end{document}